\documentstyle [twocolumn,prl,aps,floats]{revtex}  %
\begin{document}
\input epsf

\title{What Determines Inhomogeneous Line Widths in Semiconductor 
Microcavities?}

\author{D. M. Whittaker}

\address{Toshiba
Cambridge Research Centre, 260 Cambridge Science Park, Milton Road, Cambridge
CB4 4WE. United Kingdom}

\address{\em 
\ \\
\begin{minipage}[h]{6.5in}
Microcavity inhomogeneous line widths are calculated numerically using
a microscopic two dimensional model of the polariton interaction with
quantum well disorder. The calculations show show that in most
structures the line widths are determined by disorder scattering
between polariton and exciton states. This is because motional
narrowing effectively removes the contribution due to multiple
scattering between polariton states.
\\
\end{minipage}
\vspace*{-6ex}
}


\maketitle

Recent experimental results\cite{dmw-mn} have shown the importance of
motional narrowing in determining the line widths of features in the
reflectivity spectra of semiconductor microcavities. In a microcavity,
the fundamental optical excitations are polaritons, arising from the
coupling of a confined cavity photon mode to exciton states in quantum
wells within the cavity. The spectral features associated with the
polariton are broadened, both homogeneously due to the finite life
time of the cavity photon, and inhomogeneously due to the interaction
with disorder in the quantum wells. Motional narrowing occurs because
the polaritons, being part photon, have very long wavelengths compared
with those typical for excitons. The inhomogeneous broadening is
therefore substantially reduced by averaging over a large area of the
much shorter length scale disorder potential.

The important physical parameters which determine the inhomogeneous
line widths are the dispersions of the exciton and photon, the
strength of their interaction, and the statistical properties of the
disorder. A good approximation is to treat each dispersion as
parabolic, the photon mass, $M_l \sim 3 \times 10^{-5} m_e$, being a
consequence of the effects of cavity confinement\cite{dispersion}.  The
exciton-photon coupling leads to the formation of two polariton
branches, separated at resonance by the vacuum Rabi splitting, $\hbar \Omega
\sim 5$ meV, which provides a measure of the strength of the interaction.
Finally, the quantum well disorder potential is characterised by its
amplitude $V_0$, of order a few meV, and a correlation length $l_c
\sim 100$ \AA.

The original treatment of motional narrowing in Ref.\cite{dmw-mn}
assumed the weak disorder limit, $V_0 \ll \hbar \Omega$, in which only
scattering between low momentum polariton states on the same branch is
allowed. It was then possible to use a simple scaling argument to
predict the variation of the line widths with detuning. However, in a
comment on that model, Agranovich et al.\cite{agran} showed that the
same scaling argument predicts the actual value of the inhomogeneous
line width, at resonance, to be
\begin{equation}
\Gamma_{\rm 2d} \sim 
  \frac{1}{2}
  V_0^{2} 
  {\left( \frac{\hbar^2}{2 M_l l_c^2}\right)^{-1}}.
\label{eq-2d-est}
\end{equation}
For physically reasonable parameters, $\Gamma_{\rm 2d}$ is extremely
small, $\sim 10^{-4}$ meV, compared with the experimental
inhomogeneous line widths of about $0.3$ meV.

The main aim of this letter is to demonstrate that, besides the
polariton multiple scattering processes which give
Eq.(\ref{eq-2d-est}), there is a contribution to the line width due to
disorder scattering between polariton and higher momentum exciton
states. Such scattering is not included in the scaling treatment,
since it assumes the polariton dispersion is parabolic, while in fact
the two branches decouple at large wave-vectors. The more realistic
calculation described here uses a numerical simulation of a polariton
in the presence of the disorder, which automatically includes both
types of scattering.  However, in a typical structure, the small value
of the polariton multiple scattering contribution, estimated above,
means that polariton to exciton scattering processes dominate.  On the
other hand, in a high quality sample, where $V_0 < \Omega/2$,
scattering of polaritons into exciton states becomes energetically
impossible, and only the small polariton multiple scattering
contribution remains.

The numerical model used here is physically similar, though solved
differently, to the one dimensional simulations of Savona {\it et
al.}\cite{savona97}. However, motional narrowing in one dimension is
significantly less important than in two, because averaging over a 1d
strip samples much less of the variation of the potential than an
average over a 2d area of equivalent size. This difference can be seen
very clearly by calculating the 1d equivalent of Eq.(\ref{eq-2d-est}),
for the contribution of polariton multiple scattering to the line
width,
\begin{equation}
\Gamma_{\rm 1d} \sim 
  \frac{1}{2}
  {V_0^{4/3}}
  {\left( \frac{\hbar^2}{2 M_l l_c^2}\right)^{-1/3}}
\label{eq-1d-est}
\end{equation}
For the parameters that gave $\Gamma_{\rm 2d}\sim 10^{-4}$ meV,
$\Gamma_{\rm 1d} \sim 10^{-1}$ meV, three orders of magnitude greater,
and far more comparable to the experimental line widths. This comparison
suggests that the balance between polariton multiple scattering and
polariton to exciton scattering should be very different in such 1d
calculations when compared to real 2d structures.

The numerical treatment is based on the two level model of the
polariton, in which a discrete cavity mode couples to the exciton
ground state. The exciton interacts with the quantum well disorder
potential, so in-plane momentum is not conserved. Thus the model
consists of coupled two dimensional partial differential equations
describing the in-plane motion of both the exciton and cavity
photon. This model contains a number of approximations, most notably
the omission of excited exciton states and the lack of a proper
treatment of the effects of the magnetic field applied in the
experiments. These approximations restrict the possibilities for
precise comparisons with experiment, but the the good general
agreement which is obtained suggests that the essential physics is
included.

The Hamiltonian for the model system takes the form
\begin{equation}
H= \left(
   \begin{array}{cc}
      -\frac{\hbar^2}{2 M_l} \nabla^2 + \delta -i \gamma & \hbar \Omega/2  \\
      \hbar \Omega/2 & -\frac{\hbar^2}{2 M_e} \nabla^2  +
      V( \bf {r})
   \end{array}	
   \right)
\label{eq-ham}
\end{equation}
In addition to the photon mass $M_l$, and the vacuum Rabi splitting,
$\hbar \Omega$, discussed above, $M_e$ is the exciton mass, $\delta$
the detuning of the cavity photon relative to the exciton, and
$\gamma$, the photon homogeneous width (due to escape through the
mirrors).  The quantum well disorder, $V({\bf r})$, is a Gaussian
stochastic potential, constructed in the Fourier domain, following the
method described by Glutsch et al.\cite{glutsch}. Its amplitude,
$V_0$, is defined as the width at half maximum of the potential
probability distribution.

The inhomogeneously broadened spectra are calculated from the photon
Green's function 
\begin{equation}
G_k(t) = -i <\mu_{k} | e^{-iHt/\hbar} | \mu_{k}> \theta(t) 
\end{equation}
where $|\mu_k>$ represents a state of the system (not an eigenstate)
consisting of a plane wave cavity photon with in-plane wavevector $k$
and no exciton. All the results given in this letter are for normal
incidence, corresponding to $k=0$. The Green's function describes how
a photon which enters the cavity is scattered out of its initial plane
wave state by interactions with the disordered exciton.  The
plotted spectra are actually the photon spectral functions,
obtained by Fourier transforming the Green's function to give
$\tilde{G}_k(\omega)$ and extracting the imaginary part. They
correspond to what would be seen in an absorption measurement, but the
line widths should be the same in reflectivity experiments.

The Green's function is obtained using an approach similar to that of
Glutsch et al.\cite{glutsch}.  Starting from an initial state
$|\psi(0)> = |\mu_k >$, the wavefunction $|\psi(t)>$ is calculated by
numerically solving the time dependent Schr\"odinger equation using
the Hamiltonian Eq.(\ref{eq-ham}). From the solution at each time $t$,
$G_k(t)$ is found by evaluating $\mbox{$<\mu_k |\psi(t)>$} = \mbox{$<
\mu_k |e^{-iHt/\hbar} |\mu_k>$}$.  The calculation is carried out on a
two dimensional spatial grid, using a standard alternating direction
Crank-Nicholson algorithm\cite{nr}. The main difficulty is the need to
use a grid which is fine enough to reveal the correlated structure of
the potential, on a length scale \mbox{$l_c \sim 100$ \AA}, yet large
enough to avoid significant size quantisation for the low mass photon
states. The present numerical results are for a $2^{11} \times 2^{11}$
grid with spacing 100 \AA, giving photon quantisation energies $\sim
0.05$ meV, considerably smaller than typical calculated line widths of
$\sim 0.3$ meV.

\begin{figure}[tbh]
\mbox{\hspace{0.45in}
\epsfxsize=2.5in
\epsfbox{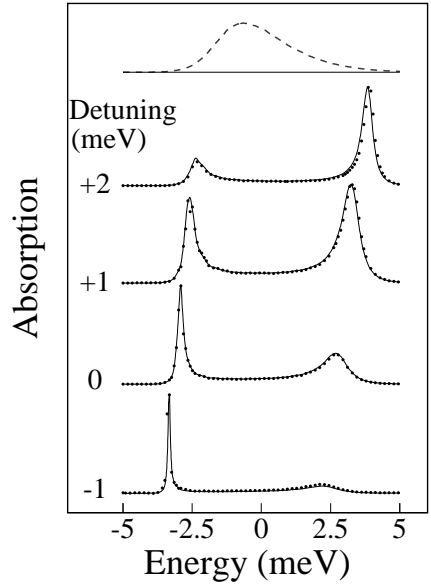} }
\vspace{-0.01in}
\caption{Numerical absorption spectra (dots) at various values of the 
detuning $\delta$. The dashed curve shows the bare exciton
lineshape. The solid lines are the predictions of the coupled
oscillator model (see text).}
\label{fig-spec}
\end{figure}

\begin{figure}[tbh]
\mbox{\hspace{0.45in}
\epsfxsize=2.5in
\epsfbox{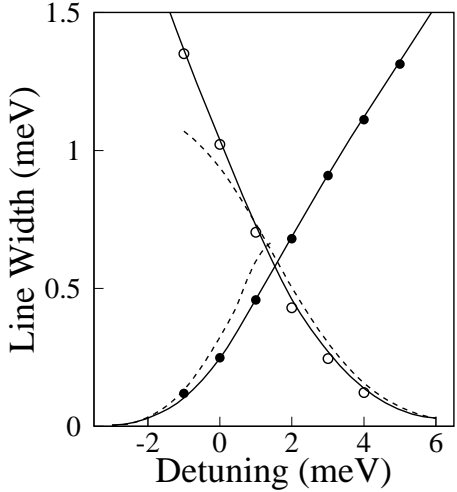} }
\vspace{0.05in}
\caption{Numerical line widths as a function of detuning, $\delta$,
for upper (open symbols) and lower (filled symbols) polaritons. The
solid lines are the predictions of the coupled oscillator model, the
dashed lines those of the absorption model.}
\label{fig-det}
\end{figure}

Fig.\ref{fig-spec} shows spectra calculated for a realistic set of
parameters: $M_l=3 \times 10^{-5} m_e$, $M_e=0.5 m_e$, $\hbar
\Omega=5$ meV, $V_0 = 3.5$ meV and $l_c =100$\AA. For clarity, the
photon homogeneous line width, $\gamma$, has been set to zero. The
spectra display the variations in intensity typical of an anticrossing
between two states, only one of which, the photon, has intrinsic
strength. Close to resonance, where the polariton branches are equal
mixtures of photon and exciton, the two peaks have similar integrated
strength.  Further away from resonance, the more photon-like peak is
strong, while the other, predominantly exciton, is weak. The figure
also shows the bare exciton spectrum, which is asymmetrically
broadened because of the finite exciton mass\cite{schnabel,glutsch-94}.

In Fig.\ref{fig-det}, the line widths measured from the spectra are
plotted as a function of detuning. The asymmetry between the line
widths of the two branches at zero detuning is very apparent: the
upper branch has a width of $\sim 1$ meV, and the lower branch 0.25
meV. When a realistic photon homogeneous broadening of $\gamma=1.25$
meV is included in the calculation, the widths become 1.4 and 0.6
meV respectively, close to the experimental values of \mbox{$\sim
1.5$} and 0.75 meV found in Ref.\onlinecite{dmw-mn}. However, because
of the lack of a proper treatment of the magnetic field used in the
experiments, this agreement can only be regarded as suggestive.

The solid lines in Figs.\ref{fig-spec}-\ref{fig-det} show the results
of a simpler model, which provides an excellent fit to the numerical
data. In this model, the full exciton Green's function
$\tilde{G}_{kk'}^e(\omega)$ is approximated by its diagonal part
$\delta_{kk'} \tilde{G}_{k}^e(\omega)$. With $k$ thus conserved, the
polariton problem simplifies to a pair of coupled oscillators, which
is easily solved to obtain the polariton Green's function
\begin{equation}
\tilde{G}_k(\omega) = 
  \frac{1}
  {\hbar \omega + i \gamma - \delta_k - (\hbar \Omega/2)^2 \,
  \tilde{G}_{k}^e(\omega) }
\label{eq-green}
\end{equation}
where $\delta_k = \delta + \hbar^2 k^2/(2 M_l)$ is the detuning for
wave-vector $k$. The bare exciton Green's functions,
$\tilde{G}_{k}^e(\omega)$ are calculated using the same type of
numerical simulation as for the full polariton. However approximate
expressions for the finite mass exciton spectral function, by Glutsch
and Bechstedt\cite{glutsch-94}, could be used for large exciton
masses.  

Eq.(\ref{eq-green}) excludes polariton multiple scattering, since the
polariton always has a definite wave-vector, $k$.  However, polariton
to exciton scattering processes are included, because the exciton
Green's function $\tilde{G}_k^e(\omega)$ takes care of the multiple
scattering of excitons between different momentum states. It has a
finite imaginary self energy determining the rate at which excitons
scatter out of the state with wave-vector $k$.  The good fit to the
numerical data which is obtained indicates that the contribution of
polariton multiple scattering is negligible for these parameters, as
the estimate of Eq.(\ref{eq-2d-est}) suggests.

In the coupled oscillator model, the difference between the widths of
the two polariton branches is a consequence of the asymmetry of the
exciton line shape, which is, in turn, a result of the finite exciton
mass. On the low energy side of the line, there is an exponential
cut-off, reflecting the distribution of minima in the disorder
potential, while on the high energy side the strength falls off more
slowly, as $\sim \omega^{-2}$, determined by the perturbation of
higher momentum plane wave states by the potential\cite{efros}. This
explanation for the asymmetry is thus similar in essence to the
suggestion by Savona {\em et al}\cite{savona97} that the larger upper
branch width is caused by scattering into higher momentum exciton
states.

An approximate expression for the polariton line widths can be found
by examining Eq.(\ref{eq-green}). For sufficiently small broadening, the
polariton lineshape approximates to a Lorentzian, with full width
$\Gamma$ dependent on the imaginary part of the denominator at the
polariton energy, $\omega_p$, according to
\begin{equation}
\Gamma= 2 |c_l|^2 \left( \gamma - \Omega^2/4 \, \mbox{Im} 
\{  \tilde{G}_{k}^e(\omega_p) \} \right)
\label{eq-abs}
\end{equation}
where $|c_l|^2$ is the photon fraction of the polariton. This
approximation is shown as the dashed lines on Fig.\ref{fig-det}. For
large photon fractions, when the polaritons are resonant with the
tails of the exciton line (see Fig.\ref{fig-spec}), the approximate
expression agrees fairly well with the exact curve. However, when
$|c_l|^2$ becomes small, Eq.(\ref{eq-abs}) predicts $\Gamma
\rightarrow 0$, while the true value tends to the exciton line width.
This behaviour can be understood more physically by noting that the
second term in Eq.(\ref{eq-abs}) describes homogeneous broadening of
the polariton due to the loss of photons in the cavity by absorption
into exciton states -- it is proportional to the photon fraction and
the exciton absorption strength
$-\mbox{Im}\{\tilde{G}_{k}^e(\omega_p)\}$.  This is to be be expected
for a state in the tail of the exciton line shape, which is too weak
to require a strong coupling treatment, and thus acts simply as a
source of absorption, like the continuum states without a magnetic
field\cite{tignon}. The absorption picture breaks down when the
polariton is resonant with the states near the centre of the exciton
line, and the full strong coupling treatment is required.

\begin{figure}[tbh]
\mbox{\hspace{0.45in}
\epsfxsize=2.5in
\epsfbox{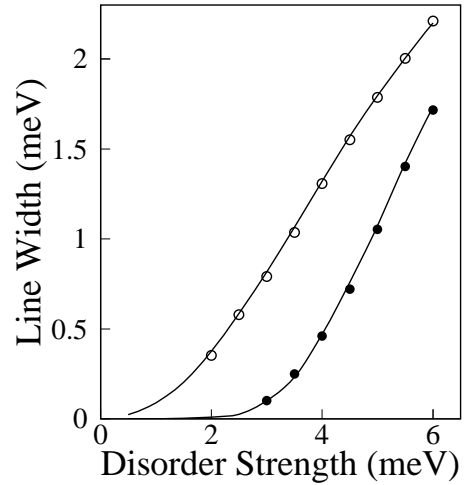} }
\vspace{0.01in}
\caption{Numerical line widths as a function of disorder strength, $V_0$,
for upper (open symbols) and lower (filled symbols) polaritons. The
solid lines are the predictions of the coupled oscillator model.}
\label{fig-pot}
\end{figure}

The coupled oscillator model described above can be shown to be a
generalisation of the phenomenological treatment of a translationally
invariant Gaussian broadened exciton, proposed by Houdr\'e {\it et
al}\cite{houdre}. The main improvement is that the Gaussian exciton
line is replaced by an asymmetic form because of the finite exciton
mass. More importantly, the present work demonstrates that motional
narrowing is the reason why such a model succeeds, in that it eliminates
the polariton multiple scattering contribution to the line width.

The most intriguing feature of the Gaussian model, discussed by Savona
and Weisbuch\cite{savona96}, is the prediction that the disorder
contribution to the line width should disappear in high quality
structures when polariton to exciton scattering becomes impossible. In
terms of the absorption picture, this happens when the polariton
states are so far into the tails of the exciton line that no
absorption occurs. The present calculations support this prediction,
as is shown by Fig.\ref{fig-pot}, where numerical and coupled
oscillator model line widths are plotted as a function of disorder
strength, $V_0$. As in Ref.\cite{savona96}, the lower branch width
rapidly becomes very small when $V_0 < \Omega/2$. The effect is much
less pronounced for the upper branch, because of the longer tail on
the high energy side of the finite mass exciton line shape.

\begin{figure}[tbh]
\mbox{\hspace{0.45in}
\epsfxsize=2.5in
\epsfbox{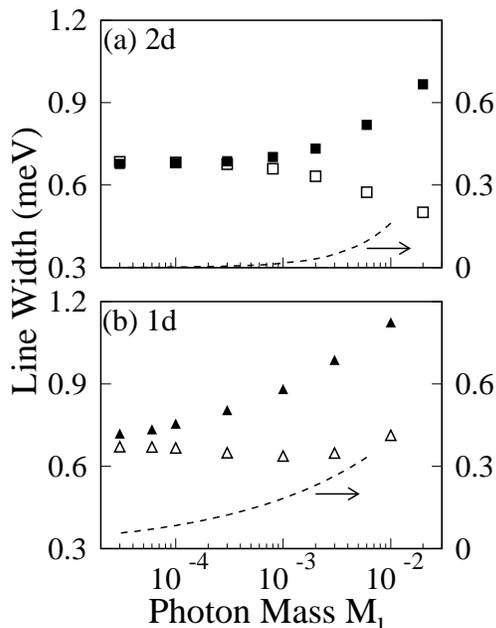} }
\vspace{0.01in}
\caption{Numerical line widths as a function of photon mass, $M_l$,
for upper (open symbols) and lower (filled symbols) polaritons
 in (a) two dimensions and (b) one dimension, with $M_e
\rightarrow \infty$. Also shown (dashed lines and right hand scale) 
are the polariton multiple scattering expressions, $\Gamma_{\rm 2d}$ and
$\Gamma_{\rm 1d}$ of Eqs. \protect{(\ref{eq-2d-est})} and 
\protect{(\ref{eq-1d-est})}.}
\label{fig-1d2d}
\end{figure}

All the numerical data in Figs.\ref{fig-spec}-\ref{fig-pot} are very
well explained by the coupled oscillator model, which includes only
polariton to exciton scattering. It is not possible in the numerical
calculations to resolve directly the very small polariton multiple
scattering term, $\Gamma_{\rm 2d}$ from Eq.(\ref{eq-2d-est}), even for
$V_0 \ll \Omega/2$, when it should be the only contribution to the
line width. However, indirect evidence for the existence of the
process can be obtained by considering non-physical situations in
which polariton multiple scattering is more important. As an example,
Fig.\ref{fig-1d2d}(a) shows the dependence on the photon mass, $M_l$,
of the numerically simulated line widths at resonance. The oscillator
model predicts that the line widths of the two branches should be
equal, as the calculation is for infinite exciton mass, and
independent of $M_l$. For the real $M_l = 3 \times 10^{-5} m_e$, the
line widths are indeed equal. However, when $M_l$ is significantly
increased a difference appears, which is clearly due to polariton
multiple scattering, as it grows in proportion to $\Gamma_{\rm 2d}$,
plotted as the dashed curve on the figure.  Fig.\ref{fig-1d2d}(b)
shows similar data for a one dimensional system. Again, the difference
between the upper and lower branch widths grows, for small $M_l$, in
proportion to $\Gamma_{\rm 1d}$\cite{width}. However, because motional
narrowing is much less effective in 1d than 2d, even for the real $M_l
= 3 \times 10^{-5} m_e$, there is a measurable difference in line
widths, some 0.05 meV. This result emphasises that there are
significant differences between the scattering processes which occur
in 1d and 2d systems.

In summary, the numerical simulations presented in this letter have
shown that there are two types of processes by which disorder
contributes to the broadening of microcavity line widths. In typical
structures, scattering between photon and exciton states
dominates. From the point of view of the photon, this is just a form
of life time broadening, caused by absorption into states in the tails
of the exciton distribution, and is thus in some senses
homogeneous. The true inhomogeneous width, the result of multiple
scattering of polaritons by the disorder, is extremely small because
of motional narrowing. However, it should be the only disorder
contribution to the width of the lower polariton branch in high
quality structures.

I would like to thank C. L. Foden, M. S. Skolnick and J. J. Baumberg
for their helpful comments on this work.


\begin{references}
\bibitem{width} The plotted width is actually half of
$\Gamma_{\rm 1d}$ in Eq. \protect{(\ref{eq-1d-est})}, as this fits
better to calculated exciton line widths in 1d.
\end{references}
\end{document}